\documentclass{aastex}
\usepackage{longtable}
\usepackage{booktabs}
\usepackage{lscape}
\usepackage{spr-astr-addons}
\usepackage{url}
\usepackage{graphicx}
\usepackage{hyperref}

\maketitle

\begin{document}

\title{Detection of wide binary and multiple nuclei of planetary nebulae using the Gaia\,DR3.}
%Wide binary central stars of planetary nebulae using Gaia DR3.}
%\slugcomment{Not to appear in Nonlearned J., 45.}
%% Running heads
\shorttitle{Detection of wide binary and multiple CSs}
\shortauthors{Ali, Khalil & Mindil}

\author{Ali, A.\altaffilmark{1}} \and
\affil{Astronomy, Space Science \& Meteorology Department, Faculty of Science, Cairo University, Giza 12613, Egypt.}
\email{afouad@sci.cu.edu.eg}
\author{Khalil, J.M. \altaffilmark{1}}
\affil{Astronomy, Space Science \& Meteorology Department, Faculty of Science, Cairo University, Giza 12613, Egypt.}
\email{Jasminesaif66@yahoo.com}
\and
\author{Mindil, A.\altaffilmark{2}}
\affil{Department of Physics, College of Science, University of Jeddah, Jeddah, Saudi Arabia.}
\email{amindil@uj.edu.sa}

%\author{fourth\altaffilmark{1}}
%\affil{Astronomy and Space Science Dept., Faculty of Science, King Abdulaziz University, 21589 Jeddah, Saudi Arabia}
%\email{\emaila}

\begin{abstract}\\
\textbf{Context}: this paper describes the detection of wide binary and multiple central stars (CSs) of Galactic planetary nebulae (PNe) using the most up-to-date data available from the Gaia Data Release 3 (Gaia\,DR3).  \\
\textbf{Aims}: the objective of this study is to benefit from the Gaia DR3's reliable measurements of parallax and proper motion to reveal the binary, ternary, and quadruple CSs of planetary nebulae. \\
\textbf{Methods}: in our search for the binary and multiple CSs in the Gaia archive, we adopted the criteria provided in the literature to ensure that the CS and its partner(s) have comparable parallaxes and proper motions. \\
\textbf{Results}: we have detected a total of 45 binary and multiple stellar systems coupled with the CSs of PNe. Based on the standard error of the parallax, this sample was divided into four categories: highest  probable, probable, possible, and uncertain systems, which comprise 18, 8, 10, and 9 objects, respectively. Except for A\,35, NGC\,246 and IC\,3568, the radial velocities of the CSs' companions are unknown for our PNe sample. The radial velocity measurements of these three companion stars are comparable to their linked CSs. The results show the detection of a highly probable quadruple system, and a probable quadruple system: NGC6853 and PHRJ1129-6012, respectively. In addition, we found one highly probable (Fr 2-42), one probable (M\,1-58) and two possible (IC 2553 and PHR\,J1123-6030) ternary CS systems. The results further show that the primary components of eight wide and very wide binary systems are actually close binary systems. Moreover, the masses of the components of nine PN binary and multiple wide systems were calculated.
\end{abstract}


%\section*{}
%\label{sec:intro}

\section{Introduction}\label{section1}
The detection of Galactic PNe, which harbor binary CSs started to grow during the last decade (see, e.g., \citet{DeMarco13}; \citet{Douchin15}; \citet{Hillwig17}; \citet{Barker18};  \citet{Boffin19}; \citet{Gonzalez20}; \citet{Aller20}); \citet{Gonzalez21}. On contrary to the identification of numerous close binary CSs, modest number of wide binary CSs is known \citet{Boffin19}. The first detected wide binary CS is the nuclei of NGC\,246 \citet{Minkowski65}. The physical pair stars have almost comparable proper motion ($\mu$) and radial velocity ($V_{rad}$) values and are separated by an angular distance ($\rho$) of 3.8\arcsec. A further photometry study was conducted by \citet{Bond99} to determine the resolved companion of NGC\,246. By fitting the suggested G8-K0 companion to the zero-age main sequence, they derived a distance of 495\,pc for the nebula. A small set of possible wide binary CSs was suggested by \citet{Cudworth73}. Of this set, the CS of NGC\,6853 was considered a very probable wide binary system, based on the proper motions of the physical pair \citet{Cudworth77}. In frame of determining the individual distances for PNe based on the spectroscopic parallax ($\pi$) method, \citet{Ciardullo99} have observed 113 PNe using the Hubble Space Telescope (HST). They identified 19 wide binary CSs, where 10, 6, and 3 of them are classified as probable, possible, and uncertain physical associations, respectively.

PNe possess not only central binary systems, but also ternary systems. A new low-mass stellar companion for the white dwarf primary component of NGC\,246 was recognized by \citet{Adam14} for the nebula to become one of the rare PNe associated with a triple system. It has also been suggested that the nebula K\,1-6 retains  a binary or ternary system \citep{Frew11}. A possible third companion for the CS of LoTr\,5 was reported by \citet{Aller19}. Furthermore \citet{Gonzalez21} reported a possible ternary system of PN Fr\,2–42. {\it As a PN can hold binary or ternary system, we show here an example of a PN (PHR\,J1129-6012) associated with a possible quaternary system}.

\citet{Soker97} has proposed four classes of interaction between the primary CS and its associated companion. (1) Very wide binaries, (2) Wide binaries, (3) Close binaries that avoid common envelopes, and (4) Close binaries that form common envelopes.  The first class has orbital periods much longer than the PN lifetime and corresponds to a physical (projected) separation ($s$) larger than 5000 au. The companions of this class can form small bubbles inside their host nebulae but these bubbles will be notable only if the companion is a red giant star. The second class has orbital periods comparable to the PN lifetime and corresponds to a physical separation of $\sim$ 100 to 1,000 au. The nebulae associated with this class show a slight deviation from axisymmetry and a deviation of the CS from the PN center. In the third class, the gravitational interactions shape the PN without the ejected material swallowing up the companion, or at least during most of the primary's mass-loss phase. In the fourth class, the companion is close enough to be engulfed by the AGB envelope.

Gaia is a space observatory operated to respond to basic inquiries regarding the composition, formation and evolution history of the Milky Way. It arrived at its space home and began to gather information in the middle of 2014. A little while later, it got ready to gather information for $\sim$ 1.7 billion sources, which represent about 1\% of the Milky Way stars. The first, second and third Gaia Data Release (Gaia\,DR1, Gaia\,DR2, and Gaia\,DR3) catalogs were published in September 2016, April 2018, and June 2022, respectively. In contrast with Gaia\,DR1 and Gaia\,DR2, Gaia\,DR3 has more precise and higher quality astrometric and photometric data as well as incorporates the measurement of new parameters.

\citet{Gonzalez20} detected nine wide binary partners linked to the CSs of eight PNe from a sample of 211 objects using accurate astrometric measurements provided by Gaia\,DR2. They limited the search to objects with relatively excellent parallax and proper motion data, with less than 30\% standard errors. Furthermore, to reduce accidental detections, they confined the search to the binary systems that have $s$ of less than 15000 au.  \citet{Gonzalez21} have created a catalog of PN central stars using Gaia EDR3, which includes astrometric and photometric data for 2035 objects. They picked 357 stars from that catalog with precise astrometric data to search for wide binary CSs. Finally, they discovered four new wide binary CS partners associated with four PNe.

There is currently no established definite limit for binary system separations \citep{Jimenez-Esteban19}. Some earlier studies, e.g., \citet{Retterer82} and \citet{Weinberg87} set the threshold at 0.1 pc (20,000 au). This limit was assumed according to the traditional theories, which believe that binary system formation occurs during the coeval fragmentation of the giant molecular cloud. Several modern studies have discovered a lot of binary systems with $s > 0 .1$  pc, and therefore two theories have been suggested to beat this issue. It was proposed by \citet{Reipurth12} that triple-star systems form very close together and can develop a hierarchical structure over the course of millions of years. One component of the system would get dynamically ejected into a remote orbit due to ejection energy gained from the collapsed orbits of the other two stars, which often causes them to appear as a single star from a distance. Thus, such weakly tied triple systems will appear to be extremely wide binaries. Another theory postulated by \citet{Kouwenhoven10} proposed that these binary and multiple systems are not primordial, but instead originate from various birth locations and become gravitationally coupled during the disintegration stage of young star clusters. There exist numerous examples of detected and confirmed wide PNe with separations $s > 0.1$ pc. HD 149414, a very wide metal-poor binary system with a separation of 0.27 pc (55000 au), was discovered by \citet{Zapatero04}. \citet{Caballero09} discovered six very wide systems with projected physical separations $s > 0.1$ pc and four of them with separations $s > 0.2$ pc. \citet{Oelkers17} claimed the discovery of the binary system with the largest separation nearly 3.2 pc. They updated their older catalogs by identifying a subset of 543 long-lived candidate binary pairs with separations between 0.002 and 1.5 pc. \citet{Galvez-Ortiz17} announced the detection of 36 low and very low-mass binary and multiple systems with separations between 200 and 92,000 au. \citet{Oh17} discovered a large number of very wide comoving pairs with significant separations $s > 1.0$ pc. Using Gaia DR2 data, \citet{Jimenez-Esteban19} developed a catalog of wide binary bright stars with  maximum separations of $s \sim 2.5$ pc.
%, and \citet{Zavada20} created a catalog of binary candidates with a separation of $s \leq 0.15$ pc.
\citet{Gonzalez23} recently performed an analysis on the widest binary systems in the Washington Double Star catalog, based on astrometric data from the Gaia DR3, and discovered three very wide systems composed of intermediate and late M dwarfs with projected physical separations of 0.33-0.41 pc.

According to \citet{Goodwin10}, binary systems are classified into four types based on the projected separation: close binaries ($s(au) < 50$); intermediate binaries ($50 < s(au) < 1000$); wide binaries ($1000 < s(au) < 10000$); and very wide (ultra-wide) binaries ($s(au) > 10000$). Wide binaries in the Galactic disc and halo can be used to restrict the formation history of the Galaxy, test the dark matter theory, and set constraints on the density of massive astrophysical compact halo objects (\citet{Kouwenhoven10} and the references therein). Recent studies have demonstrated that ultra-wide binaries are typically members of triple systems, and that close binaries frequently have a distant third companion \citep{Reipurth12} as well. For example, Proxima Centauri (Centauri C), the nearest star to the Sun, is in a very wide binary system (15000 au) with the close binary system Alpha Centauri (A) \& Alpha Centauri (B). Based on the Gaia DR2 data, \citet{El-Badry18} published a catalog including 50000 wide binaries with physical separations of $50 \leq s(au) \leq 50000$. This catalog was later enlarged to include 782017 wide binaries by relaxing the constraints on the parallax and proper motion \citep{El-Badry19}. They classified binary systems into three types. The first is a white dwarf (WD) with another WD and this type has a projected separation of $50 < s(au) < 50000$. The second is a main sequence (MS) with another MS, with $500 < s(au) < 50000$. The final type is a WD with a MS, with $s(pc) \leq 0.15$. \citet{Tian20} also used Gaia DR2 data to provide a large sample of ultra-wide binary stars with separations of $0.01 \leq s (pc) \leq 1.0$ in the solar neighbourhood.

The main objective of the present study is to enlarge the number of known wide binary central stars, aiming to quantitatively discusses their astrometric and kinematic characteristics. The physical associations of the wide binary systems analysed here are based on both the trigonometric parallaxes, which place the primary star and its binary companion at the same distance, and proper motion of each system.

Section 2 presents the data sample and the method applied for detecting the wide binary CS systems from the Gaia\,DR3 archive. The results and discussions are given in Section 3, while the conclusion is given in the last section.

\section{Searching Methodology} \label{Searching Methodology}

To classify two (three) stars as a well-defined binary or multiple system, their parallax, proper motion, and $V_{rad}$ measurements of all stars in the system must have comparable values. The Gaia project offers observations of parallax and proper motion for $\sim 1.5$ billion objects, while providing $V_{rad}$ values for only 33 million sources. The radial velocities of many PNe are known, as determined by the Doppler shift of their spectral emission lines, consequently the $V_{rad}$ of their associated CSs are also known. However, the radial velocities for almost all the CS partners are unknown. Therefore, we limited our search to binary systems where the parallax and proper motion of the pair of stars are consistent. Only the radial velocities of the three binary systems, A\,35, NGC\,246 and IC\,3568 were identified.

In the first stage of our work, we searched for wide binary systems in the sample of PNe provided by \citet{Gonzalez21} in the Gaia DR3 database. This sample consists of 2035 CSs with reliability classes (A) and (B) and 850 CSs with reliability class (C). Class (A) is defined as having colors below $-0.2$ and existing within 20\% of the PN radius, while class (B) is defined as having $f(c,d) \leq 0.5$. Class C contains all the remaining CSs, including those with unknown colors; for additional information, see \citet{Gonzalez21}.
This sample was extracted mostly from the HASH catalog \citet{Parker16} of PNe. \citet{Gonzalez21} applied the approach outlined by \citet{Gonzalez19} to reach this sample.
In addition to the previous primary source of CSs, we have selected some CSs from the literature that have been referred to as closed binary systems and possible wide binary systems, e.g., \citet{Ciardullo99}, \citet{Soker97}, \citet{Gonzalez20} , \citet{DeMarco13}, and \citet{Douchin15}. In the second stage, we rejected any stars with insufficient astrometric data and negative parallaxes. In the third stage, we constrained our search for the central star's companion to a 25\arcsec radius circle around each CS. Except for the central star's partner of A\,21 which resides at an angular separation of 23.5\arcsec, all other companions reside at angular separations less than 20\arcsec from their associated CSs. In addition, when examining the PN Fr\,2-42 in the literature, we noticed that one study \citep{Perpiny19} indicates a third companion located at an angular separation of 50\arcsec. In the final stage, the sample we obtained was subjected to a visual inspection of both the CS and its partner(s) using the Aladain Sky Atlas platform. This was done to avoid the possibility of incorrectly identifying the CS and to validate the location of its partner within the nebula.

The parallax measurements were corrected to account for the systematic zero point bias as reported by \citet{Lindegren21}. The angular separation between the binary components was converted to the projected separation using the formula $s(\textrm{au}) = \rho \times d (\textrm{pc})$, where $d$ is the distance of the binary system. As indicated in Section 1, the distribution of the projected separations between the two components of binary systems is an open question, with $s$ extending up to 0.1\,pc and 1.0\,pc for wide and very wide binary systems, respectively. Here, we adopted $50 < s\,(au) < 50000$ for wide binaries following conditions by \citet{El-Badry18} and \citet{El-Badry19}.

We grouped the predicted binary system candidates into four categories based on the parallax standard error ($\sigma_{\pi}$). The standard errors in the proper motion ($\sigma_{\mu}$) for the majority of Gaia sources belonging to these four categories are less than 20\%, with mean standard errors of $\mu_{\alpha}= 11.4\%$ and $\mu_{\delta}=8.0\%$. In general, the standard error in the parallax increases gradually with increasing the source distance. Because most PNe are far away, their parallax errors are large in comparison to their associated proper motion errors.  The four categories are as follows: (I) Highest probable binary systems: those with $\sigma_{\pi}$ less than 30\%; (II) Probable binary systems: those with $\sigma_{\pi}$ between 30\% and 50\%; (III) Possible binary systems: those with $\sigma_{\pi}$ between 50\% and 100\%; (IV) Uncertain binary systems: those with $\sigma_{\pi}$ greater than 100\%. All four categories satisfy the following constraints on proper motion and parallax proposed by El-Badry et al.(2018):
(1) $\Delta\,d - 2\,s \leq 3\sigma_{\Delta d}$ and (2) $\sigma_{\Delta \mu} < 1.5$ mas/yr, where

\begin{equation}\label{1}
 \Delta\,d = |1/\pi_{1} - 1/\pi_{2}|
\end{equation}

\begin{equation}\label{2}
\sigma_{\Delta d} = (\sigma_{\pi_{1}}^{2}/\pi_{1}^4 + \sigma_{\pi_{2}}^{2}/\pi_{1}^4)^{1/2}
\end{equation}

\begin{equation}\label{5}
\sigma_{\Delta\mu} = \frac{1}{\Delta\mu}[(\sigma_{\mu_{\alpha_{1}}^{\ast}}^{2} + \sigma_{\mu_{\alpha_{1}}^{\ast}}^{2})\Delta\mu_{\alpha}^{2}+(\sigma_{\mu_{\delta_{1}}}^{2} + \sigma_{\mu_{\delta_{2}}}^{2})\Delta\mu_{\delta}^{2}]^{1/2}
\end{equation}

\begin{equation}\label{4}
\mu_{\alpha_{i}}^{\ast} = \mu_{\alpha_{i}}\cos\delta_{i}
\end{equation}

\begin{equation}\label{3}
\Delta\,\mu = [(\mu_{\alpha_1}^{\ast} - \mu_{\alpha_2}^{\ast})^2+(\mu_{\delta_{1}}-\mu_{\delta_{1}})^2]^{1/2}
\end{equation}

\begin{equation}\label{6}
\Delta\mu_{\alpha}^{2} = (\mu_{\alpha_{1}}^{\ast} - \mu_{\alpha_{2}}^{\ast})^2
\end{equation}

\begin{equation}\label{7}
\Delta\mu_{\delta}^{2} = (\mu_{\delta_{1}}-\mu_{\delta_{2}})^2
\end{equation}

$\pi_{i}$ and $\sigma_{\pi_{i}}$ denote the parallax of the source and its standard error

\section{Results and discussions}
\subsection{The binary and multiple systems of PNe}
Table \ref{Table1}  lists our detected sample of wide binary and multiple systems. The list was divided into four categories based on the limitations and constraints explained in Section 2. Category (I) contains 18 objects; \textbf{11} of them were previously known as wide binary systems in the literature, while 7 were newly detected. Categories (II), (III), and (IV) contain 8, 10, and 9 systems, respectively, where all of them are newly detected. The table shows the PN status and morphology as collected from the HASH catalog, the reliability class as devised by \citet{Gonzalez20}, the parallax, proper motion, magnitude in G, B, R bands, and color index (B-R) of the primary CS and its partner(s). The $V_{rad}$ are listed for both wide binary components of A\,35, NGC\,246, and IC\,3568 systems and for only a few primary components, e.g., A\,33, K\,1-6, and NGC\,6853. The table also displays the calculated tangential velocity $V_t$ for each PNe.

\subsection{The close binarity of the CS of wide binary systems}
Inspecting our sample of PNe of wide binary CSs, we identified from the literature and Gaia variability identifier that eight CSs of them are close binary systems. The CS of the wide binary system NGC\,246 is a close binary of orbital period 6.84 days \citet{Aller20}. Furthermore, the CS of NGC\,5189 is a close binary with two different measured periods of 1.72 \citep{Aller20} and 4.04 days \citep{Manick15}. The CSs of  NGC 6853 \citep {De-Marco13} and NGC 6781 \citep{Douchin15} were identified as close binary systems using the I and J-band infrared flux excess (due to presence of a cold companion) technique. The CSs of NGC 6853 and NGC 6781 have close companions of M5V and M3/4 spectral types, respectively. Moreover, the CSs of A\,24, NGC 6720, PHR\,J0650+0013, PHR\,J1831-0715 nebulae are recommended as close binary systems due to their photometric variability as given in the Gaia DR3 database (Ali et al 2022, in preparation).

\subsection{Galactic populations of wide binary CSs}
We calculated the tangential velocities of the wide binary and multiple systems to show their Galactic membership. \citet{Creevey22} states that stars associated with the Galactic thin-disk have $V_{t} < 40$ km/s and those associated with the Galactic thick-disk have  $40 \leq V_{t} \leq 180$ km/s, while those associated with the Galactic halo have V$_{t} > 180$ km/s.  According to the findings in Table \ref{Table1}, 19 systems belong to the thin-disk, 17 to the thick-disk, and 2 to the Galactic halo. Therefore, it appears there is no favorite tendency for the wide binary systems of PNe to belong to any particular Galactic disk.

\begin{figure*}
\includegraphics[scale=0.50]{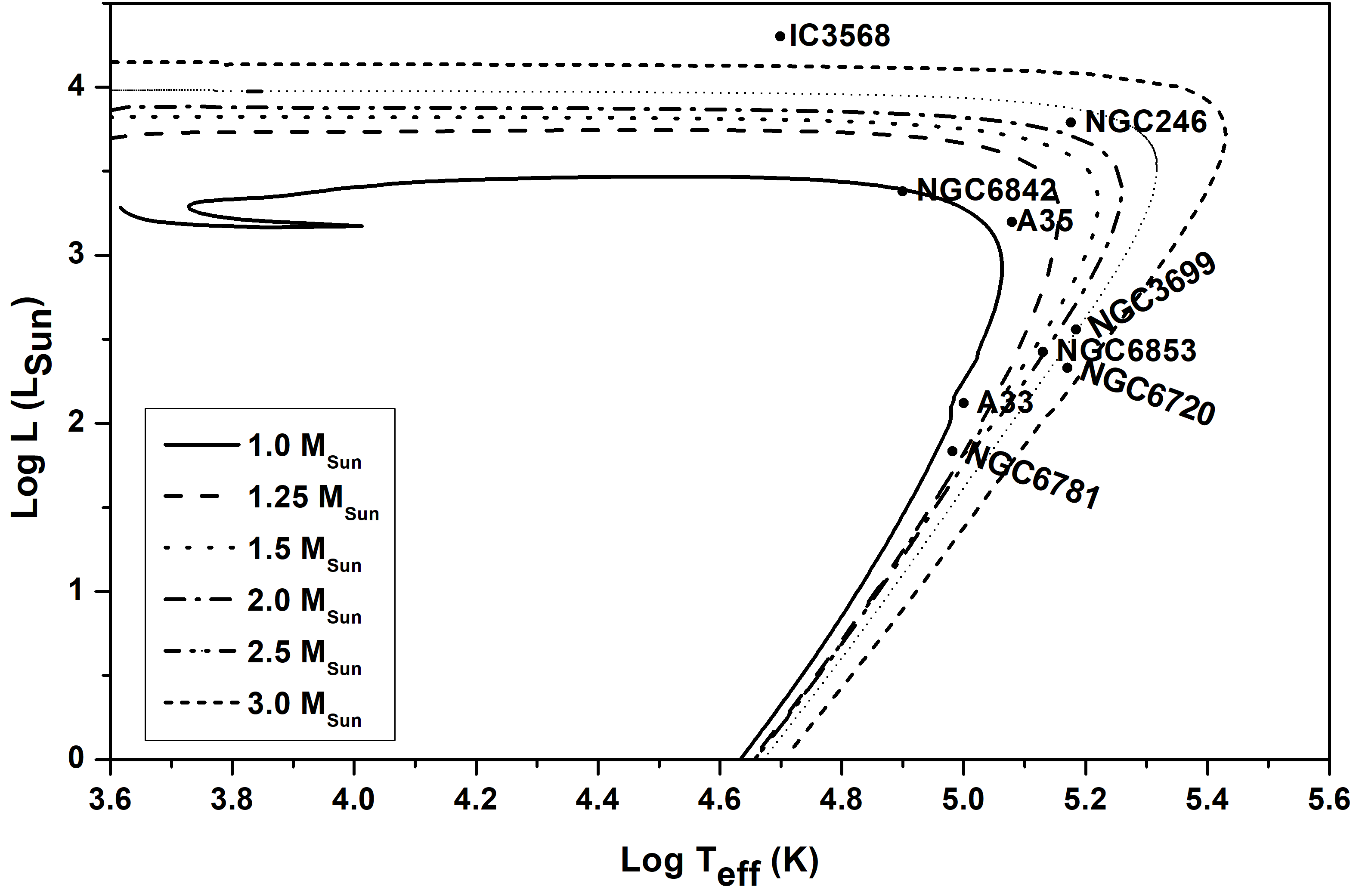}
\caption{The H–R diagram of a subset of central stars associated with wide binary and multiple system members. Mass isochrones for a solar metalicity of Z = 0.01 are displayed on the figure as taken from \citet{Miller-Bertolami16}.}
\label{Figure2}
\end{figure*}

\subsection{The masses of the wide binary components}
The effective temperature, luminosity, and surface gravity of the CSs are known for a number of PNe. As a consequence, we  used the theoretical models to calculate the initial masses of nebular progenitor stars (central stars). Figure 1 illustrates the positions of the CSs, with available data, on the H-R diagram. The CS temperatures were assembled from \citet{Weidmann20}, while the CS luminosities were computed using the Gaia distance and CS visual magnitude corrected for interstellar reddening, as described by \citet{Frew08}. The apparent visual magnitude and reddening coefficient $E(B-V)$ of the CSs were compiled from \citet{Acker92} and \citet{Frew16}, respectively. H-burning post-AGB isochrones with solar metallicity $Z=0.01$ \citep{Miller-Bertolami16} were superimposed on the H-R diagram. These tracks describe the theoretical evolution of six model stars with varying initial masses (1.0, 1.25, 1.50, 2.0, 2.5, and 3.0 M$_{\odot}$) from the beginning of the post-AGB phase to the end of the white dwarf cooling sequence phase. Table \ref{Table2} displays the CSs' initial masses of PNe plotted in Figure 1.

We computed the masses of the companion stars using the Virtual Observatory SED Analyzer (VOSA)\footnote{from the Spanish Virtual Observatory (SVO)} tool \citep{Bayo08}. In the first step, the positions, distances, and reddening coefficients of the stars were added. The coordinate and distance of the CS partner were obtained from the Gaia DR3 database. Because the Gaia's extinction parameters $A_G$ and $E(B-R)$ are not available for most objects, we used the reddening parameters $E(B-V)$ and $Av$. Here, we assumed the interstellar extinction is the same for the CS and its partner and adopted the extinction law $Av = 3.1 E(B-V)$ \citep{Fitzpatrick99}. In the second step, we compiled all of the photometric data available for the companion star in the different databases, using the VOSA interface, to construct the stellar energy distribution (SED) curve. In the third step, we determined the star's temperature and luminosity by fitting the SED curve to the convenient model. For this study, we used the BT-Settl (CIFIST) model \citep{Allard97}, which is dedicated to modelling low mass stars, brown dwarfs, and planetary mass objects. In the last step, the stellar mass was calculated using the PARSEC\,1.2 evolutionary isochrones \citep{Bressan12}, via the VOSA interface. As a result, we were able to estimate the mass, effective temperature, and luminosity of nine wide binary systems, as shown in Table \ref{Table2}. In this analysis, we limited our study to the members of Category I.

\citet{Gonzalez20} have reported the masses of the binary components of A\,24, A\,33, A\,34, NGC\,246, NGC\,3699, and NGC\,6853. Our results listed in Table \ref{Table2}, for the objects in common, are comparable with their results.
Despite the fact that \citet{Gonzalez20} have reported the mass of the A\,24 CS companion, we were unable to calculate it because we lacked the photometric information needed to create the companion's SED curve.

\subsection{Notes on some PNe hosting wide binary and multiple systems}
\begin{itemize}
  \item {\textbf{Fr\,2-42}: The HASH database listed the object as a possible PN and \citet{Perpiny19} assigned it as a triple system of three white dwarfs. \citet{Gonzalez21} classified it as a wide PN triple system. We found that the angular separation between the primary component and its first companion is 2.4\arcsec, while it is 50\arcsec between the primary and its second companion. In contrast to the conventional single blue CS, all three components of this object appear to be of blue color ($B-R < 0$). We inspected the values of the Renormalised Unit Weight Error (RUWE) factor of the three stars and found it has an unbiased value. Additionally, we looked at the B/R excess factor, which calculates the excess flux in the B and R integrated photometry relative to the G band due to the background contamination, and noticed that the three stars are of a typical small value.  A large value of this factor for a given source indicates systematic errors in the B and R photometry.  This result shows that it is unlikely that this object is a possible PN.}

  \item {\textbf{A\,35}: This object was defined as a PN by \citet{Abell66}. Later,  the object was defined as a Stromgren zone in the ambient ISM that was ionized by a hot DAO white dwarf, which has recently evolved from the PN phase \citet{Frew08}. The HASH catalog listed the object's state as an ionized ISM. According to  \citet{Grewing88}, the object has a close binary system consisting of a DAO white dwarf CS and a partner star of spectral type G8 III-IV. This companion has a projected separation of $18\pm5$ au and a suggested orbital period greater than 40 years \citep{Gatti98}. We detected a second companion at a projected separation of 1452 au,  making this object host a ternary star system. The Gaia DR3 archive provides radial velocities of $-38.1\pm4.6$ km/s for the CS and $-32.0\pm5.1$ km/s for the second companion star. The fact that the two values fall within the error range further proves the wide binarity of the CS.}

  \item {\textbf{K\,1-6}: is an asymmetric, elliptical nebula surrounding a CS of red color. \citet{Frew11} suggested that the nebula is interacting with the ISM. Moreover, they derived an a statistical distance for the object of $\sim 1.0$ kpc, however, the Gaia DR3 indicates a distance of $258\pm2$ pc for the PN central star. They also proposed that the nebula might have a binary or ternary center star.  The red color of the CS can be explained as a high interstellar extinction in the line of sight of the object or due to unresolved close binary companion. We detected another companion located 1840 au away from the PN central star.}

\item {\textbf{PHR\,J1129-6012}: According to \citet{Gonzalez20}, this nebula contains a wide binary system. Another two wide partners were identified in this work, with physical separations of 15872 au and 17862 au. Because they limited their search to stars with \textbf{$\sigma_{\pi} \leq 30\%$}, \citet{Gonzalez20} ignored both new wide companions. We noticed that the parallax and proper motion of both new stars are comparable to those of the CS, making this a probable quadruple system.}

\item {\textbf{NGC\,6853}: As explained in section 3.2, the CS of this nebula is a close binary system. We detected another two highly probable wide companions (Table 1, category I) at projected distances of 2506 and 3193 au. This makes NGC\,6853 a highly probable quadruple system, in a PN, where the CS associated with one close and two wide companions.}

\item {\textbf{M\,1-58 \& IC\,2553 \& PHR\,J1123-6030}: The CS of each of these three nebulae has two wide companions, which makes each of them a ternary system. M\,1-58 is defined as a probable ternary system with two companions at projected separations of 18054 and 22852 au, while IC\,2553 and PHR\,J1123-6030 are defined as possible ternary systems with projected separations of 29664 au, 44399 au and 38571 au, 38000 au, respectively.}

\end{itemize}

\section{Conclusions}

We used the trustworthy measurements of the parallax and proper motion, which were provided by the Gaia DR3 archive, to search for wide binary and multiple central stars in planetary nebulae. The physical association between the binary (or ternary) components of each system is primarily dependant on the consistency of their parallax and proper motion values. Following the constraints explained in Section 2, we disclosed 45-wide binary and multiple systems. Eleven of them are known in the literature. This sample was divided into four categories according to the standard error of the parallax measurements. Four of the PN sample host ternary systems (Fr\,2-42, M\,1-58, IC\,2553, PHR J1123-6030) and two host quadruple systems (NGC\,6853, PHR\,J1129-6012). We found the primary component of 8 wide binary systems are, in fact, close binary systems.(NGC\,246, NGC\,5189, NGC\,6853, NGC\,6781, A\,24, NGC\,6720, PHR\,J0650+0013, PHR\,J1831-0715), in keeping with the hypothesis of \citet{Reipurth12} . Due to the fact that the three stars of the ternary system Fr\,2–42 appear blue ($B-R < 0$), this object's classification as a possible PN in the HASH database is questioned. Unfortunately, we were unable to identify the absolute color of this stellar system since neither Gaia nor any other database provides data regarding the interstellar extinction in the line-of-sight of this ternary system. We analysed the Galactic population of this PNe sample and determined that its members do not have a preferential location in the Galaxy, with around half of them belonging to the thin-disk and the other half to the thick-disk. Finally, we succeeded in estimating the masses of nine binary and multiple systems that belong to the highest probable wide binary and multiple systems category.

\acknowledgments
The authors appreciate the anonymous referee's remarks and recommendations, which significantly enhanced this paper. This work has made use of data from the European Space Agency (ESA) mission Gaia, processed by the Gaia Data Processing and Analysis Consortium(DPAC). This research has made use of the SIMBAD database, operated at CDS, Strasbourg, France. This research has made use of "Aladin Sky Atals" developed at CDS, Strasbourg Observatory, France. This publication makes use of VOSA, developed under the Spanish Virtual Observatory (https://svo.cab.inta-csic.es) project funded by MCIN/AEI/10.13039/501100011033/ through grant PID2020-112949GB-I00.

\bibliographystyle{raa}
%\bibliography{Dist_Gaia_EDR3}
\bibliography{PaperII_PNe}
%\end{document}

\newpage
\onecolumn
\begin{landscape}
%\normalsize
\tiny	
\setlength{\tabcolsep}{1.5pt}
\renewcommand{\arraystretch}{2.5}
\setcounter{table}{0}
\begin{longtable}{llcccccllllllcccccccll}
\caption{Highest probable, probable, possible, and uncertain binary systems in planetary nebula.} \label{Table1} \\
\toprule
%\hline
     \multicolumn{4}{c}{\bf{PNe}} & reliability & \bf{Gaia DR3} & \bf{RA}  & \bf{Dec.} & \bf{Parallax} & \bf{Distance} & \multicolumn{3}{c}{\bf{Separation}} & $\mu_{\alpha}$ & $\mu_{\delta}$ & \multicolumn{4}{c}{\bf{Magnitude}} &  \bf{$V_{rad}$} & \bf{$V_{tan}$} \\  \cline{1-4} \cline{11-13} \cline{16-19}
	
$\#$   &  Name   &  Status & Shape    &  class        & ID &    (\arcdeg)  & (\arcdeg) & mas & pc & (\arcsec) & au & pc & mas/yr & mas/yr & G & B & R & B-R & km/s & km/s \\

\hline
\endhead
\multicolumn{21}{c}{\bf{Category I: Highest probable binary system}}\\
\hline		\\																																																																																																																																																																																																																																																																																																																																																																																																																																																																																																																																																																																																																																																																																																																																																																																																																																																																																																																																																																																																																																																																																																																																																																																																																																																																																																																																																																																																																																																																																																																																																																																																																																																																																																																																																																																																																																																																																																																																																																																																																																																																																																																																																																																																																																																																																																																																																																																																																																																																																																																																																																																																																																																																																																																																																																																																																																																																																																																																																																																																																																																																																																																																																																																																																																																																																																																																																																																																																																																																																																																																																																																																																																																																																																																																																																																																																																																																																																																																																																																																																																																																																																																																																																																																																																																																																																																																																																																																																																																																																																																																																																																																																																																																																																																																																																																																																																																																																																																																																																																																																																																																																																																																																																																																																																																																																																																																																																																																																																																																																																																																																																																																																																																																																																																																																																																																																																																																																																																																																																																																																																																																																																																																																																																																																																																																																																																																																																																																																																																																																																																																																																																																																																																																																																																																																																																																																																																																																																																																																																																																																																																																																																																																																																																																																																																																																																																																																																																																																																																																																																																																																																																																																																																																																																																																																																																																																																																																																																																																																																																																																																																																																																																																																																																																																																																																																																																																																																																																																																																																																																																																																																																																																																																																																																																																																																																																																																																																																																																																																																																																																																																																																																																																																																																																																																																																																																																																																																																																																																																																																																																																																																																																																																																																																																																																																																																																																																																																																																																																																																																																																																																																																																																																																																																																																																																																																																																																																																																																																																																																																																																																																																																																																																																																																																																																																																																																																																																																																																																																																																																																																																																																																																																																																																																																																																																																																																																																																																																																																																																																																																																																																																																																																																																																																																																																																																																																																																																																																																																																																																																																																																																																																																																																																																																																																																																																																																																																																																																																																																																																																																																																																																																																																																																																																																																																																																																																																																																																																																																																																																																																																																																																																																																																																																																																																																																																																																																																																																																																																																																																																																																																																																																																																																																																																																																																																																																																																																																																																																																																																																																																																																																																																																																																																																																																																																																																																																																																																																																																																																																																																																																																																																																																																																																																																																																																																																																																																																																																																																																																																																																																																																																																																																																																																																																																																																																																																																																																																																																																																																																																																																																																																																																																																																																																																																																																																																																																																																																																																																																																																																																																																																																																																																																																																																																																																																																																																																																																																																																																																																																																																																																																																																																																																																																																																																																																																																																																																																																																																																																																																																																

1	&	Fr 2-42$^{(1,2)}$	&	Possible 	&	 	&	A	&	4190499986125543168	&	298.40	&	-10.33	&	7.75	$\pm$	0.06	&	129	$\pm$	1$^*$	&	2.4	&	311	&	0.002	&	-11.5	$\pm$	0.07	&	-16.5	$\pm$	0.04	&	16.3	&	16.1	&	16.3	&	-0.2	&	 			&	12	$\pm$	0.1	\\
	&		&		&		&		&	4190499986125543296	&	298.40	&	-10.32	&	7.66	$\pm$	0.06	&	130	$\pm$	1$^*$	&		&		&		&	-10.9	$\pm$	0.07	&	-15.8	$\pm$	0.04	&	16.4	&	16.3	&	16.5	&	-0.2	&				&				\\
	&		&		&		&		&	4190500054845023488	&	298.39	&	-10.33	&	7.76	$\pm$	0.10	&	129	$\pm$	2$^*$	&	50.1	&	6444	&	0.031	&	-11.0	$\pm$	0.11	&	-16.1	$\pm$	0.06	&	17.3	&	17.3	&	17.3	&	0.0	&				&				\\
2	&	A 35	&	Ionized ISM	&	 	&		&	 3499149202247569536	&	193.39	&	-22.87	&	7.30	$\pm$	0.32	&	137	$\pm$	6$^*$	&	9.6	&	1452	&	0.007	&	-56.0	$\pm$	0.41	&	-13.1	$\pm$	0.27	&	9.4	&	9.9	&	8.7	&	1.2	&	-38	$\pm$	4.6	&	35	$\pm$	1.7	\\
	&		&		&		&		&	3499149202247569408	&	193.39	&	-22.88	&	6.01	$\pm$	0.03	&	166	$\pm$	1$^*$	&		&		&		&	-60.9	$\pm$	0.04	&	-13.4	$\pm$	0.02	&	14.3	&	15.2	&	13.3	&	1.9	&	-32	$\pm$	5.1	&				\\
3	&	A 33$^{(3)}$	&	True 	&	R	&	A	&	3827045525522912128	&	144.79	&	-2.81	&	1.00	$\pm$	0.06	&	979	$\pm$	57$^*$	&	1.8	&	1721	&	0.008	&	-14.9	$\pm$	0.06	&	9.4	$\pm$	0.05	&	15.9	&	15.7	&	15.8	&	-0.1	&	60	$\pm$	4.0	&	83	$\pm$	4.9	\\
	&		&		&		&		&	3827044765316735104	&	144.79	&	-2.81	&	1.05	$\pm$	0.06	&	933	$\pm$	49$^*$	&		&		&		&	-14.9	$\pm$	0.06	&	9.7	$\pm$	0.05	&	16.7	&	16.0	&	15.6	&	0.4	&				&				\\
4	&	K1-6 	&	True 	&	E	&	B	&	2288467186442571008	&	301.06	&	74.43	&	3.86	$\pm$	0.03	&	258	$\pm$	2$^*$	&	7.1	&	1840	&	0.009	&	-1.3	$\pm$	0.38	&	-31.4	$\pm$	0.91	&	12.3	&	12.9	&	11.5	&	1.4	&	-56	$\pm$	13.7	&	38	$\pm$	1.2	\\
	&		&		&		&		&	2288467190738512256	&	301.06	&	74.43	&	3.82	$\pm$	0.09	&	261	$\pm$	6$^*$	&		&		&		&	-2.0	$\pm$	0.10	&	-30.2	$\pm$	0.16	&	17.0	&		&		&		&				&				\\
5	&	NGC 246$^{(3)}$	&	True 	&	E	&	A	&	2376592910265354368	&	11.76	&	-11.87	&	1.80	$\pm$	0.08	&	551	$\pm$	24$^*$	&	3.7	&	1996	&	0.010	&	-16.9	$\pm$	0.10	&	-9.2	$\pm$	0.09	&	11.8	&	11.6	&	12.2	&	-0.6	&	-46	$\pm$	4.5	&	50	$\pm$	2.2	\\
	&		&		&		&		&	2376592910265354496	&	11.76	&	-11.87	&	1.88	$\pm$	0.02	&	528	$\pm$	6$^*$	&		&		&		&	-16.6	$\pm$	0.03	&	-9.0	$\pm$	0.02	&	14.2	&	14.5	&	13.6	&	0.9	&	-32	$\pm$	3.8	&				\\
6	&	NGC 6853$^{(3)}$	&	True 	&	B	&	A	&	1827256624493300096	&	299.90	&	22.72	&	2.57	$\pm$	0.04	&	387	$\pm$	6$^*$	&	6.5	&	2506	&	0.012	&	10.5	$\pm$	0.03	&	3.7	$\pm$	0.04	&	14.0	&	13.8	&	14.4	&	-0.5	&	-42	$\pm$	0.8	&	19	$\pm$	0.3	\\
	&		&		&		&		&	1827256628817680896	&	299.90	&	22.72	&	2.58	$\pm$	0.04	&	385	$\pm$	6$^*$	&	7.3	&	3193	&	0.015	&	10.3	$\pm$	0.03	&	3.8	$\pm$	0.04	&	16.1	&	17.1	&	15.2	&	2.0	&				&				\\
	&		&		&		&		&	1827256628817906432	&	299.90	&	22.72	&	2.02	$\pm$	0.36	&	490	$\pm$	88$^*$	&		&		&		&	10.1	$\pm$	0.24	&	3.0	$\pm$	0.37	&	19.5	&	19.1	&	17.8	&	1.4	&				&				\\
7	&	MeWe 2-4	&	Likely	&	E	&	A	&	 6090271154808162048	&	210.31	&	-50.67	&	0.85	$\pm$	0.11	&	1156	$\pm$	148$^*$	&	3.5	&	4012	&	0.019	&	-10.4	$\pm$	0.12	&	-0.3	$\pm$	0.10	&	17.9	&	17.8	&	18.0	&	-0.2	&	 			&	37	$\pm$	7.5	\\
	&		&		&		&		&	6090271154816460800	&	210.31	&	-50.67	&	0.86	$\pm$	0.03	&	1136	$\pm$	41$^*$	&		&		&		&	-10.5	$\pm$	0.03	&	-0.2	$\pm$	0.03	&	15.3	&	15.8	&	14.7	&	1.2	&				&				\\
8	&	NGC 6781$^{(1)}$	&	True 	&	B	&	B	&	4294123077230164736	&	289.62	&	6.54	&	2.00	$\pm$	0.07	&	495	$\pm$	17$^*$	&	8.4	&	4062	&	0.020	&	-6.9	$\pm$	0.09	&	-4.2	$\pm$	0.06	&	16.7	&	16.9	&	16.5	&	0.4	&	4	$\pm$	3.7	&	19	$\pm$	0.7	\\
	&		&		&		&		&	294123081547192960	&	289.62	&	6.54	&	2.10	$\pm$	0.15	&	472	$\pm$	35$^*$	&		&		&		&	-6.6	$\pm$	0.18	&	-3.8	$\pm$	0.15	&	18.0	&	20.3	&	16.8	&	3.5	&				&				\\
9	&	NGC 3699$^{(3)}$	&	True 	&	B	&	A	&	5336133687170599040	&	171.99	&	-59.96	&	0.73	$\pm$	0.08	&	1333	$\pm$	148$^*$	&	3.22	&	4619	&	0.022	&	-3.2	$\pm$	0.08	&	1.1	$\pm$	0.08	&	17.6	&	17.6	&	17.7	&	-0.1	&	-16	$\pm$	7.8	&	13	$\pm$	2.5	\\
	&		&		&		&		&	5336133687204227840	&	171.99	&	-59.96	&	0.63	$\pm$	0.04	&	1536	$\pm$	101$^*$	&		&		&		&	-3.3	$\pm$	0.04	&	1.1	$\pm$	0.04	&	16.5	&	17.1	&	15.8	&	1.2	&				&				\\
10	&	PN G030.8+03.4a$^{(1)}$	&	Possible	&	S	&	B	&	4272532521447280384	&	278.88	&	-0.26	&	1.87	$\pm$	0.22	&	529	$\pm$	62$^*$	&	9.9	&	4896	&	0.024	&	2.0	$\pm$	0.21	&	-8.8	$\pm$	0.17	&	17.7	&	18.6	&	16.5	&	2.1	&	 			&	23	$\pm$	2.7	\\
	&		&		&		&		&	4272532517149931008	&	278.88	&	-0.26	&	2.16	$\pm$	0.37	&	460	$\pm$	78$^*$	&		&		&		&	2.7	$\pm$	0.34	&	-8.5	$\pm$	0.29	&	19.2	&	21.0	&	17.7	&	3.3	&				&				\\
11	&	VSP 2-18	&	True 	&	R	&	C	&	4104350432587192320	&	276.81	&	-14.14	&	0.40	$\pm$	0.08	&	2393	$\pm$	461$^*$	&	2.6	&	5345	&	0.026	&	-2.4	$\pm$	0.08	&	-4.5	$\pm$	0.07	&	16.8	&	17.6	&	15.7	&	2.0	&	 			&	60	$\pm$	11.6	\\
	&		&		&		&		&	4104350398227452032	&	276.80	&	-14.14	&	0.57	$\pm$	0.16	&	1718	$\pm$	494$^*$	&		&		&		&	-2.3	$\pm$	0.20	&	-4.3	$\pm$	0.16	&	18.3	&	19.5	&	16.9	&	2.6	&				&				\\
12	&	NGC 6842	&	True 	&	R	&	A	&	2028777318992611072	&	298.76	&	29.29	&	0.46	$\pm$	0.04	&	2113	$\pm$	176$^*$	&	3.1	&	6877	&	0.033	&	-3.7	$\pm$	0.03	&	-8.8	$\pm$	0.04	&	16.1	&	16.2	&	15.8	&	0.4	&	-5	$\pm$	40.0	&	98	$\pm$	8.3	\\
	&		&		&		&		&	 2028777318992610816	&	298.76	&	29.29	&	0.41	$\pm$	0.12	&	2323	$\pm$	671$^*$	&		&		&		&	-3.8	$\pm$	0.10	&	-8.4	$\pm$	0.13	&	18.1	&	18.7	&	17.2	&	1.5	&				&				\\
13	&	A 24$^{(3)}$	&	True 	&	B	&	A	&	3088991026757468800	&	117.91	&	3.01	&	1.39	$\pm$	0.10	&	710	$\pm$	49$^*$	&	13.5	&	9913	&	0.048	&	-4.3	$\pm$	0.10	&	-0.7	$\pm$	0.06	&	17.4	&	17.2	&	17.8	&	-0.6	&	13	$\pm$	4.0	&	15	$\pm$	1.1	\\
	&		&		&		&		&	3088990923678255744	&	117.91	&	3.00	&	1.30	$\pm$	0.04	&	758	$\pm$	24$^*$	&		&		&		&	-4.4	$\pm$	0.04	&	-1.1	$\pm$	0.03	&	15.6	&	16.1	&	15.0	&	1.1	&				&				\\
14	&	A 34$^{(3)}$	&	True 	&	R	&	A	&	5690534730341025408	&	146.40	&	-13.17	&	0.85	$\pm$	0.07	&	1160	$\pm$	90$^*$	&	9.1	&	10555	&	0.051	&	3.2	$\pm$	0.06	&	-9.2	$\pm$	0.06	&	16.4	&	16.2	&	16.8	&	-0.6	&	 		 	&	54	$\pm$	4.3	\\
	&		&		&		&		&	5690534734636923520	&	146.40	&	-13.17	&	0.84	$\pm$	0.03	&	1160	$\pm$	36$^*$	&		&		&		&	3.3	$\pm$	0.02	&	-9.2	$\pm$	0.02	&	14.8	&	15.2	&	14.3	&	0.9	&				&				\\
15	&	SB 36$^{(3)}$	&	Likely	&	 	&		&	4036463835977400064	&	268.59	&	-39.18	&	0.68	$\pm$	0.03	&	1434	$\pm$	61$^*$	&	11.1	&	14945	&	0.072	&	4.2	$\pm$	0.03	&	-4.7	$\pm$	0.03	&	15.0	&	15.5	&	14.3	&	1.2	&	35	$\pm$	15.0	&	40	$\pm$	1.9	\\
	&		&		&		&		&	4036463801649079808	&	268.58	&	-39.18	&	0.78	$\pm$	0.03	&	1258	$\pm$	48$^*$	&		&		&		&	4.3	$\pm$	0.03	&	-4.7	$\pm$	0.02	&	14.6	&	15.0	&	14.1	&	0.9	&				&				\\
16	&	NGC 6720$^{(1)}$	&	True 	&	E	&	A	&	2090486618786534784	&	283.40	&	33.03	&	1.27	$\pm$	0.04	&	777	$\pm$	27$^*$	&	18.5	&	16134	&	0.078	&	1.6	$\pm$	0.04	&	2.5	$\pm$	0.05	&	15.6	&	15.6	&	16.1	&	-0.5	&	63	$\pm$	14.0	&	11	$\pm$	0.4	\\
	&		&		&		&		&	2090486687506009472	&	283.39	&	33.03	&	1.02	$\pm$	0.22	&	967	$\pm$	206$^*$	&		&		&		&	1.2	$\pm$	0.23	&	2.6	$\pm$	0.23	&	18.8	&	16.1	&	16.0	&	0.0	&				&				\\
17	&	IC 3568	&	True 	&	R	&	A	&	 1720138697015345024	&	188.28	&	82.56	&	0.44	$\pm$	0.04	&	2266	$\pm$	189$^*$	&	14.9	&	39604	&	0.192	&	1.5	$\pm$	0.05	&	-2.4	$\pm$	0.04	&	12.9	&	12.4	&	12.7	&	-0.3	&	-41	$\pm$	0.8	&	26	$\pm$	2.6	\\
	&		&		&		&		&	1720138666951697536	&	188.25	&	82.56	&	0.33	$\pm$	0.02	&	3050	$\pm$	174$^*$	&		&		&		&	1.5	$\pm$	0.02	&	-2.3	$\pm$	0.02	&	13.0	&	13.6	&	12.2	&	1.4	&	-47	$\pm$	1.5	&				\\
18	&	Vy 1-4	&	True 	&	R	&	A	&	4253934209252777344	&	283.51	&	-6.44	&	0.17	$\pm$	0.05	&	5260	$\pm$	1437$^*$	&	9.7	&	49327	&	0.239	&	-5.7	$\pm$	0.06	&	-8.1	$\pm$	0.04	&	15.7	&	14.3	&	14.5	&	-0.2	&	55	$\pm$	20.7	&	271	$\pm$	74.3	\\
	&		&		&		&		&	4253934213542598272	&	283.51	&	-6.44	&	0.19	$\pm$	0.04	&	4911	$\pm$	1164$^*$	&		&		&		&	-3.7	$\pm$	0.05	&	-8.5	$\pm$	0.04	&	15.7	&	16.6	&	14.8	&	1.8	&				&				\\

\hline	
\multicolumn{21}{c}{\bf{Category II: Probable binary system}}\\
\hline \\
19	&	PHR J1129-6012$^{(3)}$	&	True 	&	E	&	B	&	5335989479389919104	&	172.46	&	-60.20	&	0.34	$\pm$	0.06	&	2763	$\pm$	504$^*$	&	6.2	&	15872	&	0.077	&	-7.0	$\pm$	0.07	&	2.3	$\pm$	0.06	&	17.2	&	17.6	&	16.5	&	1.1	&	 			&	57	$\pm$	18.4	\\
	&		&		&		&		&	5335989479346740096	&	172.46	&	-60.20	&	0.41	$\pm$	0.17	&	2357	$\pm$	1001	&		&		&		&	-6.6	$\pm$	0.19	&	2.7	$\pm$	0.16	&	18.9	&	18.8	&	17.6	&	1.1	&				&				\\
	&		&		&		&		&	5335989479389918336	&	172.46	&	-60.20	&	0.34	$\pm$	0.15	&	2819	$\pm$	1279	&	4.9	&	14126	&	0.068	&	-7.3	$\pm$	0.15	&	1.8	$\pm$	0.14	&	18.6	&		&		&		&				&				\\
	&		&		&		&		&	5335989479389919360	&	172.46	&	-60.20	&	0.32	$\pm$	0.07	&	3003	$\pm$	632$^*$	&	3.8	&	9554	&	0.046	&	-7.0	$\pm$	0.07	&	2.7	$\pm$	0.06	&	17.1	&	17.3	&	16.3	&	1.1	&				&				\\
20	&	Hen 2-207	&	True 	&	B	&	B	&	5952713419747721600	&	259.89	&	-45.89	&	1.44	$\pm$	0.35	&	688	$\pm$	167$^*$	&	14.8	&	10411	&	0.050	&	-1.9	$\pm$	0.44	&	-1.6	$\pm$	0.30	&	19.4	&	19.9	&	18.3	&	1.6	&	-36	$\pm$	5.7	&	7	$\pm$	2.4	\\
	&		&		&		&		&	5952713419747758080	&	259.89	&	-45.88	&	1.37	$\pm$	0.51	&	719	$\pm$	268	&		&		&		&	-1.5	$\pm$	0.64	&	-2.3	$\pm$	0.42	&	19.9	&	20.9	&	19.4	&	1.6	&				&				\\
21	&	PHR J0650+0013	&	True 	&	B	&	B	&	3113542949606809088	&	102.67	&	0.23	&	0.15	$\pm$	0.03	&	5858	$\pm$	1196$^*$	&	3.5	&	19161	&	0.093	&	-0.4	$\pm$	0.04	&	0.7	$\pm$	0.04	&	15.2	&	15.5	&	1.0	&	1.1	&				&	25	$\pm$	5.4	\\
	&		&		&		&		&	3113542949610601216	&	102.67	&	0.23	&	0.18	$\pm$	0.06	&	5091	$\pm$	1817	&		&		&		&	-0.3	$\pm$	0.10	&	0.6	$\pm$	0.12	&	16.5	&	16.7	&	0.9	&	0.9	&	 		 	&				\\
22	&	M 1-58	&	True 	&	R	&	A	&	 4107013823419786624	&	280.74	&	-11.11	&	0.47	$\pm$	0.29	&	2056	$\pm$	957	&	10.2	&	22852	&	0.111	&	-0.8	$\pm$	0.32	&	-4.4	$\pm$	0.28	&	18.7	&	15.5	&	15.0	&	0.5	&	57	$\pm$	9.7	&	45	$\pm$	27.9	\\
	&		&		&		&		&	 4107013827713941632	&	280.74	&	-11.12	&	0.40	$\pm$	0.18	&	2424	$\pm$	717$^*$	&		&		&		&	-0.5	$\pm$	0.21	&	-4.4	$\pm$	0.18	&	18.2	&	19.0	&	17.4	&	1.6	&				&				\\
	&		&		&		&		&	 4107013827714017536	&	280.74	&	-11.11	&	0.45	$\pm$	0.13	&	2142	$\pm$	633$^*$	&	8.6	&	18054	&	0.088	&	-1.3	$\pm$	0.15	&	-4.2	$\pm$	0.14	&	18.1	&	18.9	&	17.2	&	1.6	&				&				\\
23	&	NGC 5189	&	True 	&	B	&	A	&	5863702868275424384	&	203.39	&	-65.97	&	0.68	$\pm$	0.02	&	1433	$\pm$	39$^*$	&	14.2	&	24585	&	0.119	&	-12.7	$\pm$	0.02	&	-3.3	$\pm$	0.02	&	14.5	&	14.4	&	14.3	&	0.0	&	-10	$\pm$	4.0	&	43	$\pm$	2.5	\\
	&		&		&		&		&	5863702863954968192	&	203.40	&	-65.98	&	0.48	$\pm$	0.17	&	2030	$\pm$	705	&		&		&		&	-12.0	$\pm$	0.15	&	-4.4	$\pm$	0.18	&	18.9	&	18.8	&	17.7	&	1.1	&				&				\\
24	&	Hf 4	&	True 	&	B	&	A	&	5258145052514126848	&	153.90	&	-58.85	&	0.32	$\pm$	0.16	&	2960	$\pm$	1463	&	9.4	&	27718	&	0.134	&	-7.6	$\pm$	0.20	&	7.5	$\pm$	0.17	&	18.8	&	19.3	&	17.5	&	1.8	&	22	$\pm$	12.0	&	125	$\pm$	77.9	\\
	&		&		&		&		&	5258145048215455488	&	153.90	&	-58.85	&	0.32	$\pm$	0.06	&	2938	$\pm$	539$^*$	&		&		&		&	-8.0	$\pm$	0.07	&	6.3	$\pm$	0.06	&	17.2	&	17.6	&	16.2	&	1.4	&				&				\\
25	&	DSH J2003.3+2135 	&	True 	&	B	&	A	&	 1823929193070538624	&	300.84	&	21.60	&	0.33	$\pm$	0.16	&	2848	$\pm$	1405	&	14.8	&	40401	&	0.196	&	-5.5	$\pm$	0.10	&	-6.3	$\pm$	0.15	&	18.4	&	18.4	&	18.5	&	0.0	&	 		 	&	115	$\pm$	58.5	\\
	&		&		&		&		&	1823929197392350976	&	300.84	&	21.60	&	0.37	$\pm$	0.12	&	2611	$\pm$	875	&		&		&		&	-5.5	$\pm$	0.07	&	-6.5	$\pm$	0.11	&	17.9	&	18.5	&	17.1	&	1.3	&				&				\\
26	&	PN PB 2	&	True 	&	S	&	B	&	5516695417937365376	&	125.17	&	-46.38	&	0.25	$\pm$	0.07	&	3737	$\pm$	985$^*$	&	14.2	&	48603	&	0.236	&	-1.8	$\pm$	0.08	&	2.1	$\pm$	0.08	&	15.8	&	14.1	&	13.7	&	0.4	&	 		 	&	46	$\pm$	13.7	\\
	&		&		&		&		&	5516695482362147712	&	125.16	&	-46.38	&	0.30	$\pm$	0.12	&	3108	$\pm$	1194	&		&		&		&	-1.9	$\pm$	0.14	&	2.5	$\pm$	0.15	&	18.5	&	18.9	&	18.0	&	0.9	&				&				\\

\hline	
\multicolumn{21}{c}{\bf{Category III: Possible binary system}}\\
\hline \\	
27	&	PN G059.1-01.4	&	True 	&	R	&	C	&	 1827775185956173824	&	296.87	&	22.47	&	1.04	$\pm$	1.00	&	946	$\pm$	907	&	3.55	&	2889	&	0.014	&	-2.5	$\pm$	1.13	&	-4.1	$\pm$	1.33	&	20.7	&	21.5	&	19.7	&	1.8	&	 			&	21	$\pm$	21.7	\\
	&		&		&		&		&	 1827775151622403840	&	296.87	&	22.47	&	1.45	$\pm$	0.97	&	681	$\pm$	456	&		&		&		&	-2.3	$\pm$	0.90	&	-4.1	$\pm$	1.48	&	20.6	&	21.3	&	19.7	&	1.6	&				&				\\
28	&	NGC 2899	&	True 	&	B	&	A	&	 5307241785737966080	&	141.76	&	-56.11	&	0.45	$\pm$	0.26	&	2134	$\pm$	1210	&	6.3	&	12542	&	0.061	&	-2.3	$\pm$	0.50	&	-3.0	$\pm$	0.44	&	19.3	&		&		&		&	3	$\pm$	2.8	&	34	$\pm$	23.1	\\
	&		&		&		&		&	 5307241785737968256	&	141.76	&	-56.11	&	0.52	$\pm$	0.03	&	1848	$\pm$	99$^*$	&		&		&		&	-2.7	$\pm$	0.03	&	-2.4	$\pm$	0.03	&	15.7	&	16.2	&	14.9	&	1.2	&				&				\\
29	&	PHR J0911-4205	&	True 	&	E	&	B	&	 5427382290279252736	&	137.95	&	-42.09	&	0.85	$\pm$	0.35	&	1154	$\pm$	472	&	10.8	&	13902	&	0.067	&	-3.1	$\pm$	0.34	&	3.0	$\pm$	0.36	&	19.9	&	20.6	&	19.0	&	1.7	&	 		 	&	21	$\pm$	10.0	\\
	&		&		&		&		&	 5427382294568898432	&	137.95	&	-42.09	&	0.69	$\pm$	0.38	&	1420	$\pm$	783	&		&		&		&	-3.6	$\pm$	0.34	&	2.5	$\pm$	0.39	&	20.0	&	20.5	&	19.1	&	1.4	&				&				\\
30	&	A 21	&	True 	&	E	&	A	&	 3163546505053645056	&	112.26	&	13.25	&	1.69	$\pm$	0.07	&	586	$\pm$	24$^*$	&	23.5	&	17680	&	0.086	&	-2.9	$\pm$	0.07	&	-8.5	$\pm$	0.06	&	15.9	&	15.8	&	16.3	&	-0.5	&	29	$\pm$	5.2	&	25	$\pm$	1.0	\\
	&		&		&		&		&	3163546470692680192	&	112.26	&	13.24	&	1.07	$\pm$	0.63	&	918	$\pm$	539	&		&		&		&	-3.0	$\pm$	0.64	&	-8.8	$\pm$	0.54	&	20.1	&	21.4	&	19.0	&	2.4	&				&				\\
31	&	PHR J1831-0715	&	Possible 	&	I	&	C	&	4159811654519010304	&	277.82	&	-7.26	&	0.63	$\pm$	0.53	&	1551	$\pm$	1307	&	14	&	20796	&	0.101	&	-4.1	$\pm$	0.68	&	-5.1	$\pm$	0.59	&	19.8	&	21.0	&	18.6	&	2.4	&	 		 	&	49	$\pm$	41.9	\\
	&		&		&		&		&	 4159811650232150784	&	277.82	&	-7.26	&	0.69	$\pm$	0.51	&	1420	$\pm$	1058	&		&		&		&	-3.3	$\pm$	0.57	&	-5.5	$\pm$	0.48	&		&		&	18.2	&		&				&				\\
32	&	Hf 69	&	True 	&	E	&	A	&	 5333350548355207168	&	175.41	&	-62.48	&	0.43	$\pm$	0.25	&	2237	$\pm$	1289	&	10.1	&	21423	&	0.104	&	-6.5	$\pm$	0.27	&	1.3	$\pm$	0.24	&	19.4	&	19.3	&	19.6	&	-0.3	&	 		 	&	36	$\pm$	42.2	\\
	&		&		&		&		&	 5333350544047527552	&	175.41	&	-62.48	&	0.48	$\pm$	0.32	&	2005	$\pm$	1341	&		&		&		&	-6.4	$\pm$	0.36	&	1.0	$\pm$	0.32	&	19.6	&	21.0	&	18.1	&	3.0	&				&				\\
33	&	Pa 159	&	Possible 	&		&	B	&	 4211173239978256128	&	288.78	&	-6.18	&	0.44	$\pm$	0.27	&	2165	$\pm$	1338	&	12.1	&	28222	&	0.137	&	6.4	$\pm$	0.28	&	-1.5	$\pm$	0.26	&	19.1	&	19.9	&	18.2	&	1.6	&				&	69	$\pm$	43.1	\\
	&		&		&		&		&	 4211172866310572544	&	288.78	&	-6.18	&	0.38	$\pm$	0.10	&	2500	$\pm$	659$^*$	&		&		&		&	5.1	$\pm$	0.15	&	-1.7	$\pm$	0.15	&	17.5	&	18.1	&	16.7	&	1.4	&				&				\\
34	&	IC 2553	&	True 	&	B	&	A	&	 5252805514840414592	&	152.34	&	-62.61	&	0.27	$\pm$	0.12	&	3514	$\pm$	1626	&	8.6	&	29664	&	0.144	&	-4.8	$\pm$	0.15	&	2.4	$\pm$	0.13	&	16.2	&	12.6	&	12.9	&	-0.3	&	37	$\pm$	6.3	&	58	$\pm$	44.3	\\
	&		&		&		&		&	5252805480451026048	&	152.33	&	-62.62	&	0.28	$\pm$	0.23	&	3385	$\pm$	2753	&		&		&		&	-5.0	$\pm$	0.26	&	1.8	$\pm$	0.23	&	19.3	&	18.6	&	18.3	&	0.3	&				&				\\
	&		&		&		&		&	 5252805514810842496	&	152.34	&	-62.61	&	0.29	$\pm$	0.21	&	3214	$\pm$	2309	&	13.2	&	44399	&	0.215	&	-4.0	$\pm$	0.25	&	3.4	$\pm$	0.22	&	19.2	&	19.5	&	18.8	&	0.8	&				&				\\
35	&	PN G062.0+02.5	&	True 	&	R	&	A	&	2025098146936552576	&	294.62	&	26.96	&	0.15	$\pm$	0.11	&	6143	$\pm$	4586	&	5.78	&	32805	&	0.159	&	-2.4	$\pm$	0.09	&	-4.6	$\pm$	0.13	&	18.2	&	17.9	&	17.3	&	0.5	&	 			&	164	$\pm$	125.3	\\
	&		&		&		&		&	 2025098146936551168	&	294.62	&	26.97	&	0.17	$\pm$	0.06	&	5209	$\pm$	1856	&		&		&		&	-2.7	$\pm$	0.05	&	-4.8	$\pm$	0.07	&	17.3	&	18.3	&	16.3	&	2.0	&				&				\\
36	&	PHR J1123-6030	&	Possible 	&	I	&	A	&	 5337515429678393984	&	170.96	&	-60.51	&	0.33	$\pm$	0.02	&	2848	$\pm$	159$^*$	&	14.1	&	38571	&	0.187	&	-5.0	$\pm$	0.02	&	2.2	$\pm$	0.02	&	14.3	&	14.6	&	13.8	&	0.8	&	 			&	47	$\pm$	4.3	\\
	&		&		&		&		&	5337515429678384640	&	170.96	&	-60.51	&	0.36	$\pm$	0.32	&	2623	$\pm$	2316	&		&		&		&	-5.4	$\pm$	0.39	&	1.7	$\pm$	0.28	&	19.7	&	20.6	&	18.7	&	1.9	&				&				\\
	&		&		&		&		&	 5337515429678387200	&	170.96	&	-60.51	&	0.32	$\pm$	0.28	&	2998	$\pm$	2669	&	13	&	38000	&	0.184	&	-4.7	$\pm$	0.34	&	1.4	$\pm$	0.31	&	19.2	&	19.3	&	17.9	&	1.4	&				&				\\

\hline	
\multicolumn{21}{c}{\bf Category IV: Uncertain binary system}\\
\hline \\	
37	&	JaSt 95	&	True 	&	R	&	B	&	4057385514989465216	&	268.40	&	-28.48	&	0.64	$\pm$	0.25	&	1527	$\pm$	591	&	7	&	10472	&	0.051	&	-4.2	$\pm$	0.25	&	-4.6	$\pm$	0.14	&	18.2	&	19.3	&	16.5	&	2.8	&	 			&	44	$\pm$	17.9	\\
	&		&		&		&		&	4057373660945752320	&	268.40	&	-28.48	&	0.67	$\pm$	0.78	&	1465	$\pm$	1709	&		&		&		&	-4.1	$\pm$	0.64	&	-5.1	$\pm$	0.38	&	19.1	&	20.3	&	17.6	&	2.7	&				&				\\
38	&	Pa54	&	Likely 	&		&	A	&	 4303586921715070592	&	297.60	&	11.04	&	0.54	$\pm$	0.70	&	1793	$\pm$	2330	&	6.1	&	10920	&	0.053	&	-2.5	$\pm$	0.71	&	-5.0	$\pm$	0.51	&	19.9	&	20.0	&	20.1	&	-0.1	&				&	49	$\pm$	64.3	\\
	&		&		&		&		&	 4303586921713477376	&	297.60	&	11.04	&	0.54	$\pm$	0.74	&	1788	$\pm$	2448	&		&		&		&	-1.3	$\pm$	0.72	&	-4.8	$\pm$	0.59	&	20.1	&	21.2	&	19.2	&	1.9	&				&				\\
39	&	PN G048.0-04.4	&	True 	&	R	&	B	&	4315045722705910528	&	293.88	&	11.35	&	0.21	$\pm$	0.28	&	4464	$\pm$	5994	&	3.5	&	15606	&	0.076	&	-3.2	$\pm$	0.25	&	-6.1	$\pm$	0.22	&	19.2	&	18.9	&	18.6	&	0.3	&	 			&	156	$\pm$	210.1	\\
	&		&		&		&		&	 4315045722659102720	&	293.89	&	11.35	&	0.21	$\pm$	0.17	&	4453	$\pm$	3719	&		&		&		&	-4.1	$\pm$	0.15	&	-6.5	$\pm$	0.15	&	18.5	&	19.1	&	17.9	&	1.2	&				&				\\
40	&	BMP J1827-0611	&	Possible 	&	E	&	C	&	 4160190470657792384	&	276.82	&	-6.20	&	0.25	$\pm$	0.11	&	3737	$\pm$	1678	&	5.4	&	19607	&	0.095	&	-3.3	$\pm$	0.13	&	-7.1	$\pm$	0.10	&	17.5	&	19.6	&	16.1	&	3.5	&	 			&	147	$\pm$	66.1	\\
	&		&		&		&		&	 4160190470657780736	&	276.82	&	-6.20	&	0.27	$\pm$	0.64	&	3525	$\pm$	8474	&		&		&		&	-4.2	$\pm$	0.90	&	-7.0	$\pm$	0.59	&	19.9	&	21.1	&	18.6	&	2.5	&				&				\\
41	&	PN A66 23 	&	True 	&	R	&	A	&	 5588562590306924416	&	115.83	&	-34.75	&	0.17	$\pm$	0.62	&	5426	$\pm$	20220	&	6.7	&	33846	&	0.164	&	-3.5	$\pm$	0.60	&	3.7	$\pm$	0.72	&	20.5	&	20.5	&	20.0	&	0.5	&				&	132	$\pm$	532.8	\\
	&		&		&		&		&	 5588562586007420032	&	115.82	&	-34.75	&	0.20	$\pm$	0.17	&	4677	$\pm$	4111	&		&		&		&	-3.2	$\pm$	0.16	&	3.9	$\pm$	0.18	&	19.0	&	19.9	&	18.1	&	1.9	&				&				\\
42	&	M 1-73	&	True 	&	B	&	B	&	 4318420669299100288	&	295.29	&	14.95	&	0.22	$\pm$	0.03	&	4186	$\pm$	483$^*$	&	9.3	&	35462	&	0.172	&	-1.8	$\pm$	0.02	&	-4.2	$\pm$	0.02	&	14.5	&	13.8	&	13.1	&	0.7	&	-14	$\pm$	11.5	&	98	$\pm$	11.3	\\
	&		&		&		&		&	 4318420673599214848	&	295.29	&	14.95	&	0.27	$\pm$	0.52	&	3440	$\pm$	6591	&		&		&		&	-1.4	$\pm$	0.46	&	-4.7	$\pm$	0.41	&	19.9	&	20.3	&	19.0	&	1.3	&				&				\\
43	&	JaSt2-9	&	True 	&	E	&	B	&	 4056567852115947136	&	268.20	&	-29.29	&	0.38	$\pm$	1.16	&	2547	$\pm$	825	&	15.1	&	35004	&	0.170	&	-5.7	$\pm$	1.20	&	-6.9	$\pm$	0.71	&	20.0	&		&		&		&	 			&	108	$\pm$	349.9	\\
	&		&		&		&		&	 4056567856394396672	&	268.20	&	-29.29	&	0.46	$\pm$	0.39	&	2090	$\pm$	1749	&		&		&		&	-6.9	$\pm$	0.49	&	-7.4	$\pm$	0.33	&	19.1	&	18.3	&	16.9	&	1.4	&				&				\\
44	&	IRAS 18252-0512	&	True 	&	S	&	B	&	4257140836167961984	&	276.99	&	-5.17	&	0.12	$\pm$	0.28	&	7175	$\pm$	16408	&	5.7	&	38213	&	0.185	&	-3.0	$\pm$	0.33	&	-4.6	$\pm$	0.28	&	18.7	&	18.4	&	16.4	&	2.0	&	 			&	213	$\pm$	487.2	\\
	&		&		&		&		&	4257140797480655360	&	276.99	&	-5.17	&	0.14	$\pm$	0.80	&	6233	$\pm$	34654	&		&		&		&	-3.1	$\pm$	0.99	&	-5.8	$\pm$	0.87	&	20.1	&	21.9	&	18.8	&	3.1	&				&				\\
45	&	PB 10	&	True 	&	B	&	B	&	4315832144050466304	&	292.06	&	12.33	&	0.23	$\pm$	0.40	&	4003	$\pm$	6950	&	14.3	&	52308	&	0.254	&	-1.8	$\pm$	0.31	&	-5.7	$\pm$	0.31	&	19.0	&	16.9	&	15.8	&	1.1	&	 			&	121	$\pm$	210.2	\\
	&		&		&		&		&	4315832526337103872	&	292.06	&	12.33	&	0.28	$\pm$	0.08	&	3313	$\pm$	932$^*$	&		&		&		&	-2.2	$\pm$	0.07	&	-5.8	$\pm$	0.06	&	17.1	&	18.6	&	15.9	&	2.8	&				&				\\

\multicolumn{21}{l}{References:(1)\citet{Gonzalez21}, \citet{Perpiny19}, \citet{Gonzalez20}}\\
\multicolumn{21}{l}{The symbol ($^*$) indicates that the star is of distance error less than 30\%.}\\
\hline \\
\label{tab:my-table}\\
\end{longtable}
\end{landscape}

%\end{document}
\begin{table}
\tiny
\centering
\caption{The temperature, luminosity and mass of a subset of wide binary planetary nebulae.} \label{Table2}
\scalebox{1.1}{
\begin{tabular}{llcccc}
\hline \\
 PN	&	Gaia DR3$\_$ID &	$E(B-V)$	&	$Log\,T_{eff}$ (K)	&	$Log\,L$ ($L_{\odot}$)	&	Mass $M_{\odot}$ \\
\hline
A\,35 (CS)	&	 3499149202247569536	&	0.26	&	5.08	&	3.20	&	1.10	\\
A\,35 (Companion)	&	3499149202247569408	&		&	3.59	&	-1.14	&	0.60	\\
A\,33 (CS)	&	3827045525522912128	&	0.03	&	5.00	&	2.12	&	1.05	\\
A\,33 (Companion)	&	3827044765316735104	&		&	3.65	&	-0.48	&	0.89	\\
NGC\,246 (CS)	&	2376592910265354368	&	0.02	&	5.18	&	3.79	&	2.25	\\
NGC\,246 (Companion)	&	2376592910265354496	&		&	3.73	&	-0.31	&	0.89	\\
NGC\,6853 (CS)	&	1827256624493300096	&	0.05	&	5.13	&	2.43	&	2.00	\\
NGC\,6853 (Companion 1)	&	1827256628817680896	&		&	3.59	&	-1.15	&	0.60	\\
NGC\,6853 (Companion 2)	&	1827256628817906432	&		&	6.52	&	-1.85	&	0.35	\\
NGC\,6781 (CS)	&	4294123077230164736	&	0.58	&	4.98	&	1.83	&	1.20	\\
NGC\,6781 (Companion)	& 4294123081547192960	&		&	3.58	&	-1.19	&	0.55	\\
NGC\,3699 (CS)	&	5336133687170599040	&	0.31	&	5.18	&	2.56	&	2.50	\\
NGC\,3699(Companion)	&	5336133687204227840	&		&	3.74	&	-0.08	&	0.99 	\\
NGC\,6842 (CS)	&	2028777318992611072	&	0.45	&	4.90	&	3.38	&	1.00	\\
NGC\,6842 (Companion)	&	 2028777318992610816	&		&	3.62	&	-0.12	&	0.74	\\
NGC\,6720 (CS)	&	2090486618786534784	&	0.04	&	5.17	&	2.33	&	2.75	\\
NGC\,6720 (Companion)	&	2090486687506009472	&		&	3.52	&	-0.81	&	0.41	\\
IC\,3568 (CS)	&	 1720138697015345024	&	0.12	&	4.70	&	3.44	&	1.00	\\
IC\,3568 (Companion)	&	1720138666951697536	&		&	3.65	&	1.85	&	1.77	\\

\hline
\end{tabular}}
\end{table}
\end{document}